\magnification=\magstep1
\voffset6truemm
\nopagenumbers

\def\II{{\rm 1\!\hskip-1pt I}}
\font\gross=cmbx12 scaled \magstep2

\font\pl=cmssq8 scaled \magstep1
\font\sc=cmcsc10
\font\sft=cmss10 scaled \magstep1

\headline={\ifnum\pageno=1\hfill\else
{\it I.G. Avramidi and G. Esposito:
Universal Functions in Euclidean Quantum Gravity}
\hfill\rm\folio\fi}

{\null
\vskip-1.5cm
\hskip5cm{ \hrulefill }
\vskip-.55cm
\hskip5cm{ \hrulefill }
\smallskip
\vskip1mm
\hskip5cm{{\pl \ University of Greifswald (February, 1997)}} 
\smallskip
\hskip5cm{ \hrulefill }
\vskip-.55cm
\hskip5cm{ \hrulefill } 
\bigskip 
\hskip5cm{\ hep-th/9702150}
\bigskip
\smallskip
\hskip5cm{\ Submitted to:}
\medskip
\hskip5cm{\ \sft Gravity Research Foundation}
\smallskip
\hskip5cm{\ \sft Essay Contest}
\vfill
\centerline {\gross Universal Functions}
\bigskip
\centerline {\gross in Euclidean Quantum Gravity}
\bigskip
\centerline {{\sc Ivan G Avramidi}$^{1,}$
\footnote{$^*$}{On leave of absence from Research Institute
for Physics, Rostov State University, Stachki 194,
344104 Rostov-on-Don, Russia. E-mail: 
avramidi@rz.uni-greifswald.de}
{\sc and Giampiero Esposito}$^{2,3}$}
\bigskip
\noindent
\centerline
{\it ${ }^{1}$Department of Mathematics, University
of Greifswald,}
\centerline{\it  Jahnstr. 15a, 17487 Greifswald, Germany}
\medskip
\centerline
{\it ${ }^{2}$Istituto Nazionale di Fisica Nucleare,
Sezione di Napoli,}
\centerline{\it Mostra d'Oltremare, Padiglione 20,
80125 Napoli, Italy}
\medskip
\noindent
\centerline
{\it ${ }^{3}$Dipartimento di Scienze Fisiche,
Universit\`a degli Studi di Napoli ``Federico II",}
\centerline{\it  Mostra d'Oltremare,
Padiglione 19, 80125 Napoli, Italy}
\bigskip
\vfill
\noindent
{\bf Abstract}. A key problem in the attempt to quantize the
gravitational field is the choice of boundary conditions. These
are mixed, in that spatial and normal components of metric
perturbations obey different sets of boundary conditions. 
In the covariant quantization scheme
this leads to a boundary operator involving both normal and
tangential derivatives of metric perturbations. On studying the
corresponding heat-kernel asymptotics, one finds that
universal, tensorial, nonpolynomial structures contribute 
through the integrals over the boundary of linear combinations of all
geometric invariants of the problem. 
These universal functions
are independent of conformal rescalings of the background
metric, and they might lead to a deep revolution in the
current understanding of quantum gravity.

\vfill
\eject
\baselineskip=18pt

\noindent
The division of physics into differential equations for the
fields, and boundary conditions for the solutions of such
equations, has proved to be very useful both in cosmology
and in the current understanding of (quantized) field theories.
In particular, in the sum-over-histories formulation of
quantum gravity, an important task is to give a precise
definition of the $\langle {\rm out} \mid {\rm in} \rangle$ 
amplitudes of going from suitable in-data on
an initial spacelike three-surface,
to suitable out-data on a final spacelike three-surface. 
One can then try to perform a semiclassical
evaluation of the path-integral. 
Despite the well known
lack of perturbative renormalizability of a quantum gravity
theory based on the Einstein's action, one can actually discover
a lot of new exciting properties even just from a careful
analysis of the one-loop semiclassical approximation. Our
essay is devoted to this problem for pure gravity in four
dimensions, when a compact Riemannian four-manifold, $(M,g)$, 
with boundary, $\partial M$, is studied. 
Although no further assumption is
made, the reader should be aware that our work corresponds
to the Euclidean approach to quantum cosmology, rather than
to the analysis of $\langle {\rm out} \mid {\rm in} \rangle$
amplitudes with the associated asymptotic regions.

To begin, let us assume that spatial components of metric
perturbations, say $h_{ij}$, are set to zero at the boundary:
$$
\Bigr[h_{ij}\Bigr]_{\partial M}=0
\; \; \; \; .
\eqno (1)
$$
Of course, this is suggested by what one can do in linearized
theory at the classical level. A basic ingredient in our
analysis is that (1) should be preserved under infinitesimal
diffeomorphisms on metric perturbations
(here, $\nabla$ is the Levi-Civita
connection of the background):
$$
{ }^{\varphi}h_{ab}=h_{ab}+\nabla_{(a} \; \varphi_{b)}\ \ .
\eqno (2)
$$
Their action on
$h_{ij}$ reads
$$
{ }^{\varphi}h_{ij}=h_{ij}
+\widehat\nabla_{(i}\varphi_{j)}
+K_{ij} \varphi_{0}
\; \; \; \; ,
\eqno (3)
$$
where $\varphi_{b}$ is the ghost one-form, $K_{ij}$
is the extrinsic-curvature tensor of $\partial M$, and 
$\widehat\nabla$ denotes three-dimensional covariant differentiation
tangentially with respect to the Levi-Civita connection of
the boundary. It is then clear that, if $K_{ij}$ does not
vanish, a necessary and sufficient condition for the
preservation of the boundary conditions (1) under the
transformations (3) is that the whole ghost one-form vanishes
on the boundary, i.e.
$$
\Bigr[\varphi_{a}\Bigr]_{\partial M}=0
\; \; \; \; .
\eqno (4) 
$$
At this stage, the remaining set of boundary conditions on
metric perturbations, whose invariance under infinitesimal
diffeomorphisms (2) is guaranteed by (4), involves setting to
zero at the boundary a linear gauge-averaging functional 
in the Faddeev-Popov scheme:
$$
\Bigr[\Phi_{a}(h)\Bigr]_{\partial M}=0\ \ .
\eqno (5)
$$
What happens is that, under the 
transformations (2), one finds
$$
\Phi_{a}({ }^{\varphi}h)=\Phi_{a}(h)
+{\cal F}_{a}^{\; \; b} \; \varphi_{b}
\; \; \; \; ,
\eqno (6)
$$
where ${\cal F}_{a}^{\; \; b}$ is the ghost operator.
By expanding then the ghost one-form into a complete 
orthonormal set of eigenfunctions of ${\cal F}_{a}^{\; \; b}$
with Dirichlet boundary conditions (4), one sees that
the boundary conditions (5) are invariant under (2).

In particular, if a covariant gauge-averaging functional
of the de Donder type is used, i.e.
$$
\Phi_{a}(h) \equiv \nabla^{b} \Bigr(h_{ab}-{1\over 2}
g_{ab} g^{cd}h_{cd} \Bigr)
\; \; \; \; ,
\eqno (7)
$$
the boundary conditions (5) include both normal and tangential
derivatives of the normal components $h_{00}$
and $h_{0i}$.
It is then possible to express both (1) and
(5) by a single equation
$$
\Bigr[{\cal B}h \Bigr]_{\partial M}=0
\; \; \; \; ,
\eqno (8)
$$
where $\cal B$ is the boundary operator, defined as
($\II$ denotes $\delta_{\; \; (a}^{c} \; \delta_{\; \; b)}^{d}$)
$$
{\cal B} \equiv(\II-\Pi)\Bigr(H \nabla_{n}
+\Gamma^{i}{\widehat \nabla}_{i}+S \Bigr)
+\mu \Pi
\; \; \; \; .
\eqno (9)
$$
With our notation, $n^{b}$ denotes the normal to 
$\partial M$, $\nabla_{n} \equiv n^{a} 
\nabla_{a}$ is the normal derivative, $\mu$ is a 
dimensional parameter, and $q, \Pi, \Gamma^{i}$ and
$S$ are tensors defined by 
$$
q_{ab} \equiv g_{ab}-n_{a}n_{b} \ \ ,
\eqno (10)
$$
$$
\Pi_{ab}^{\; \; \; cd} \equiv q_{\; \; (a}^{c} \;
q_{\; \; b)}^{d}
\; \; \; \; , 
\eqno (11)
$$
$$
H^{ab \; cd} \equiv g^{a(c} \; g^{d)b}
-{1\over 2}g^{ab}g^{cd}
\; \; \; \; ,
\eqno (12)
$$
$$
\Gamma_{\; \; ab}^{i \; \; \; \; \; cd} \equiv
n_{a}n_{b}e^{i(c} \; n^{d)}
-n_{(a} \; e_{\; \; b)}^{i} \; n^{c} n^{d}
\; \; \; \; ,
\eqno (13)
$$ 
$$
S_{ab}^{\; \; \; cd} \equiv -n_{a}n_{b}n^{c}n^{d}
+2n_{(a} \; e_{\; \; b)}^{i} \; e^{j(c} \; n^{d)}
\Bigr[K_{ij}-\gamma_{ij}({\rm tr}K)\Bigr]
\; \; \; \; ,
\eqno (14)
$$
where $e_a^{i}$ is a local basis of one-forms on the boundary,
and
$
\gamma^{ij}=g^{ab} \; e_{\; \; a}^{i} \; e_{\; \; b}^{j}
$
is the induced metric on the boundary.

The form (9) of the boundary operator is not generic, but
depends, as we said, on the choice (7) for the 
gauge-averaging functional, jointly with (1) and (5). We
are thus studying just one of the possible schemes for mixed
boundary conditions in Euclidean quantum gravity, following 
the work in Refs. [1--6]. Our form of the boundary operator
has been also obtained in Ref. [7], within the framework of
Becchi-Rouet-Stora-Tyutin invariant boundary conditions in
quantum field theory. Note that the matrix 
$\Gamma^{2}\equiv \gamma_{ij}\Gamma^i\Gamma^j$, 
which, from (13), reads [6]
$$
\Gamma^{2}=-{3\over 2}n_{a}n_{b}n^{c}n^{d}
-n_{(a} \; q_{\; \; b)}^{(c} \; n^{d)}
\; \; \; \; ,
\eqno (15)
$$
commutes with $S$:
$$
\Gamma^{2}S=S \Gamma^{2}={3\over 2}n_{a}n_{b}n^{c}n^{d}
-n_{(a} \; e_{\; \; b)}^{i} \;
e^{j(c} \; n^{d)} \Bigr[K_{ij}-\gamma_{ij}({\rm tr}K)\Bigr]
\; \; \; \; .
\eqno (16)
$$
However, $\Gamma^{2}$ does not commute with the matrices 
$\Gamma^{i}$. Indeed, the explicit calculation shows that
$$
\Gamma^{2}\Gamma^{i}=-{3\over 2}n_{a}n_{b}
e^{i(c} \; n^{d)}+{1\over 2}n_{(a} \; e_{\; \; b)}^{i}
n^{c}n^{d}
\; \; \; \; ,
\eqno (17)
$$
whereas
$$
\Gamma^{i}\Gamma^{2}=-{1\over 2}n_{a}n_{b}e^{i(c} \; n^{d)}
+{3\over 2}n_{(a} \; e_{\; \; b)}^{i} n^{c}n^{d}
\; \; \; \; .
\eqno (18)
$$
These remarks are of crucial importance for the following
reasons. A similar form of the boundary operator (9), when $\Pi$ 
does not occur and $H=\II$, i.e.
$$
{\cal B}=\nabla_{n}
+\Gamma^{i}{\widehat \nabla}_{i}
+{1\over 2}({\widehat \nabla}_{i}\Gamma^{i})
+S
\; \; \; \; ,
\eqno (19)
$$
for a general second-order operator $P$ of Laplace type 
($\nabla$ being a connection and $E$ some 
endomorphism of a vector bundle $V$):
$$
P \equiv -g^{ab}\nabla_{a}\nabla_{b}-E
\; \; \; \; ,
\eqno (20)
$$
was studied in Ref. [2].
In general, the matrices $\Gamma^{i}$ and $S$ do not have the form 
(13) and (14) but satisfy the conditions 
$\Gamma^{i\dag}=-\Gamma^i$, $S^{\dag}=S$ [2,6]. 

In the corresponding 
asymptotic expansion of the integrated heat-kernel [6], it
is convenient to introduce a smooth function on $M$, say $f$,
which makes it possible to recover the distributional 
behaviour of the heat-kernel near the boundary. One then
finds in four dimensions, as $t \rightarrow 
0^{+}$, the asymptotics
$$
{\rm Tr}_{L^{2}} \Bigr(f e^{-tP} \Bigr) \sim
(4\pi t)^{-2} \sum_{n=0}^{\infty}t^{n/2}
a_{n/2}(f,P)
\; \; \; \; ,
\eqno (21)
$$
where the coefficients $a_{n/2}(f,P)$ are obtained
by integrating geometric invariants over $M$ (interior
terms) and $\partial M$ (boundary terms). 
More precisely,
the consideration of all the invariants which can be built
from the operators $P$ (20) and ${\cal B}$ (19)
shows that the first two coefficients in (21) can be written
in the form [2,6]
$$
a_{0}(f,P)=\int_{M}{\rm Tr}(f)
\; \; \; \; ,
\eqno (22)
$$
$$
a_{1/2}(f,P)= \int_{\partial M}{\rm Tr}(\rho(\Gamma) f)
\; \; \; \; .
\eqno (23)
$$

The occurrence of $\Gamma^{i}$ in the boundary operator 
(19) leads to many additional invariants, further to the
standard contributions for Dirichlet or Neumann boundary
conditions, in the general formulae for 
higher-order heat-kernel coefficients $a_{n/2}(f,P)$.
In particular, when $\Gamma^{2}$ 
commutes with $\Gamma^{i}$ and $S$, one can 
control all additional contributions. 
The coefficient $a_{1}$ has then the form [2,6]
$$
\eqalignno{
\; & a_{1}(f,P)= \int_{M}{\rm Tr} \Bigr[f(\alpha_{1}E
+\alpha_{2}R_{\; \; a}^{a})\Bigr]\cr&
+\int_{\partial M}{\rm Tr}\Bigr[f\Bigr(
b_{0}(\Gamma)({\rm tr}K)
+\sigma_{1}(\Gamma)K_{ij}\Gamma^{i}\Gamma^{j}
+b_{2}(\Gamma)S\Bigr)
+b_{1}(\Gamma)\nabla_{n} f\Bigr]
\; \; \; \; .
&(24)\cr}
$$
Remarkably, while the parameters 
$\alpha_{1}$ and $\alpha_{2}$ are universal
constants, $\rho(\Gamma)$, $\sigma_{1}(\Gamma)$, 
$b_{0}(\Gamma)$, $b_{1}(\Gamma)$ and
$b_{2}(\Gamma)$ turn out to be {\it universal functions}. 
This means
that they are functions of position on the boundary, and their
dependence on $\Gamma^{i}$ is realized through analytic 
functions of $\Gamma^{2}$. Thus,
by construction, these functions are independent of conformal
rescalings of the background metric. Moreover, all parameters 
in (24) are also independent of the dimension of $M$.
Further (assuming for simplicity $\widehat\nabla_i\Gamma^j=0$),
to obtain the form of
$a_{3/2}(f,P)$, one has to consider all 
possible contractions of the matrices 
$\Gamma^{i}$ (together with the normal $n^a$ and the metric $\gamma^{ij}$) 
with geometric objects that can be put symbolically in the form
$$
fKK\; \; , \; \; 
fKS \; \; , \; \; 
f {\widehat \nabla}K\; \; , \; \; 
f {\widehat \nabla}S \; \; , \; \; 
f R\; \; , \; \; fF
$$
and $K\nabla_n f$ (here, $R$ and $F$ 
denote the Riemann curvature
of $M$, and the curvature of the bundle connection, respectively).
Similarly, the general form of $a_{2}(f,P)$ receives contributions
from all contractions of $\Gamma^{i}$, $n^a$ and $\gamma^{ij}$ 
with geometric terms of the form: 
$$
fKKK\; , \; f KKS \; , \; 
f KSS\; , \;
f RK\; , \; f FK\; , \; f E K\; , \; 
f RS
\; , \; f FS \; , 
$$
$$
f K{\widehat \nabla}K\; , \; 
f S {\widehat \nabla} K
\; , \; f K{\widehat \nabla} S \; , \; 
f S {\widehat \nabla} S \; , \; 
f {\widehat \nabla} {\widehat \nabla} K\; ,
$$
$$
f {\widehat \nabla} {\widehat \nabla} S \; , \;
f \nabla R\; , \; f \nabla F\; , \;
f \nabla E \; ,
$$
as well as
$$
KK\nabla_n f \; , \; 
KS\nabla_n f \; , \;
({\widehat \nabla}K) \nabla_n f \; , \;
({\widehat \nabla}S) \nabla_n f \; , \;
R\nabla_n f \; , \; 
F\nabla_n f
$$
and $K\nabla_n\nabla_n f$. 

In our essay, however, we are concerned with Euclidean 
quantum gravity in four dimensions (four dimensions being
more relevant for the world we live in). Thus, focusing on
our original problem, we have to point out that a more basic
difficulty occurs. 

In the simple case considered in Refs. [2] and [6], when 
the matrices $\Gamma^2$ and $\Gamma^i$ commute, 
all tensor structures of rank $m$ 
built from $\Gamma^i$ have the form
$$
T^{i_1\cdots i_m}(\Gamma^j)
=T(\Gamma^2)\Gamma^{i_1}\cdots\Gamma^{i_m}
\; \; \; \; ,
\eqno (25)
$$
where $T(\Gamma^2)$ are universal 
functions of $\Gamma^2$ only.
As we have seen, this is not the case, however, for the gravitational
field, since the matrices $\Gamma^2$ and $\Gamma^i$ do not commute
(see (17) and (18)).
This means that the tensor structures $T^{i_1\cdots i_m}(\Gamma^j)$ 
on the boundary, formed by the
matrices $\Gamma^i$, do not have the simple form (25). 
In particular, even the scalar functions, 
$\alpha(\Gamma^j)$, cannot be always 
presented as the trace of functions of $\Gamma^2$ 
only, and the second-rank tensors $T^{ij}(\Gamma^l)$ 
are not polynomial in $\Gamma^i$. For example, 
there are infinitely many different tensors of the type
$$
T^{ij}_{(m)}(\Gamma^l)
\equiv{\rm Tr}\,\{\alpha_{(m)}(\Gamma^2)\Gamma^{i}
\beta_{(m)}(\Gamma^{2})\Gamma^{j}\}
\; \; \; \; ,
$$
which can contribute already to $a_{1}$. 
Thus, for the generalized boundary operator (9) 
with arbitrary noncommuting matrices $\Gamma^i$, 
even the coefficient $a_{1}$ is unknown.

One thus faces a highly nontrivial problem. On one hand, analytic
results exist for the $a_{2}$ coefficient with boundary operator
(9) in the particular case of a flat Euclidean background 
bounded by a three-sphere [3,5]. Moreover, it has been shown in
Ref. [4] that the boundary operator (9) leads to a 
self-adjoint boundary-value problem on metric perturbations. 
However, in the noncommuting case relevant for gravity,
even the building blocks of geometric invariants involving 
$\Gamma^{i}$ are unknown. That is why it remains unclear
how to write a general and unambiguous formula
for heat-kernel coefficients. The solution of this problem is of
the greatest importance in quantum gravity for the following
reasons:
\vskip 0.3cm
\noindent
(i) to improve the understanding of BRST invariant boundary
conditions [7];
\vskip 0.3cm
\noindent
(ii) to obtain an entirely geometric description of one-loop
divergences in quantum gravity and quantum supergravity [5];
\vskip 0.3cm
\noindent
(iii) as a first step towards the quantization in arbitrary
gauges on manifolds with boundary;
\vskip 0.3cm
\noindent
(iv) to clarify the differences between Yang-Mills fields
and the gravitational field;
\vskip 0.3cm
\noindent
(v) to complete the application of the effective-action
programme to perturbative quantum gravity.

Further to this, we would like to end by emphasizing that 
the problem remains of how to build and use a
non-perturbative theory of the effective action in quantum
gravity. This investigation, jointly with the analysis
of universal functions of the quantized gravitational field,
would lead to an entirely new vision, showing how deep is the
impact of boundary conditions and effective-action methods
on the attempt to combine the ideas of quantum physics and
general relativity.
\medskip
\noindent
The work of I.G.A. was supported by the Deutsche
Forschungsgemeinschaft.
\vskip1cm
\leftline {\bf References}
\vglue1cm
\item {[1]}
Barvinsky A. O. (1987) {\it Phys. Lett.} {\bf B 195}, 344.
\item {[2]}
McAvity D. M. and Osborn H. (1991) {\it Class. Quantum
Grav.} {\bf 8}, 1445.
\item {[3]}
Esposito G. , Kamenshchik A. Yu. , Mishakov I. V. and
Pollifrone G. (1995) {\it Phys. Rev.} {\bf D 52}, 3457.
\item {[4]}
Avramidi I. G. , Esposito G. and Kamenshchik A. Yu. (1996)
{\it Class. Quantum Grav.} {\bf 13}, 2361.
\item {[5]}
Esposito G. , Kamenshchik A. Yu. and Pollifrone G. (1997)
{\it Euclidean Quantum Gravity on Manifolds with Boundary}
(Dordrecht: Kluwer).
\item {[6]}
Avramidi I. G. and Esposito G. (1997) {\it Heat-Kernel
Asymptotics with Generalized Boundary Conditions}
(hep-th/9701018).
\item {[7]}
Moss I. G. and Silva P. J. (1997) {\it Phys. Rev.}
{\bf D 55}, 1072.

\bye